\documentclass[%
 reprint,
 showpacs,
 showkeys,
 preprintnumbers,
 amsmath,amssymb,
 aps,
 prl,
  longbibliography,
 ]{revtex4-1}

\usepackage[breaklinks=true,colorlinks=true,anchorcolor=blue,citecolor=blue,filecolor=blue,menucolor=blue,pagecolor=blue,urlcolor=blue,linkcolor=blue]{hyperref}
\usepackage{graphicx}% Include figure files
\usepackage{xcolor}

\begin{document}

\title{Haunted quantum contextuality versus value indefiniteness -- a minority report}

\author{Karl Svozil}
\affiliation{Institute of Theoretical Physics, Vienna
    University of Technology, Wiedner Hauptstra\ss e 8-10/136, A-1040
    Vienna, Austria}
\email{svozil@tuwien.ac.at} \homepage[]{http://tph.tuwien.ac.at/~svozil}

\date{\today}

\begin{abstract}
Physical entities are ultimately (re)constructed from elementary yes/no events, in particular clicks in detectors or measurement devices recording quanta. Recently, the interpretation of certain such clicks has given rise to unfounded claims which are neither necessary nor sufficient, although they are presented in that way. In particular, clicks can neither inductively support nor ``(dis)prove'' the Kochen-Specker theorem, which is a formal result that has a deductive proof by contradiction. More importantly, the alleged empirical evidence of quantum contextuality, which is ``inferred'' from violations of bounds of classical probabilities by quantum correlations, is based on highly nontrivial assumptions, in particular on physical omniscience.
\end{abstract}

%\preprint{CDMTCS preprint nr. 372/2009}
\maketitle

Time and again, in coffee houses and elsewhere,
members of the Viennese experimental physics community reminded me
always to keep in mind that all our physical ``facts''
are ultimately derived and constructed from detector clicks.
It is this basic wisdom that, when consequentially applied to recent experiments,
suggests to rethink certain claims of empirical proof.

Let us, for the sake of properly assessing the situation,
review some historical cornerstones.
Motivated by certain, as it turned out inapplicable,
no-go theorems by von Neumann regarding hidden parameters,
Bell came forward with criteria for classical probabilities and expectations,
resembling the {\em conditions of possible experience} that
had been contemplated by Boole a century earlier~\cite{Pit-94}.
Essentially, these criteria state that,
if one forces the (counterfactual) physical co-existence
upon certain finite sets of complementary, incompatible, potential observables
--
meaning that every single one could be measured,
although due to complementarity
it is impossible to simultaneously measure all of them
--
the associated potential measurement outcomes are
subject to certain algebraic bounds.

As these probabilistic bounds are not satisfied by quantum observables,
the respective measurements outcomes cannot consistently co-exist~\cite{peres222};
at least not under the classical
presumptions entering the calculations leading to these bounds.
These arguments have subsequently been strengthened by the Kochen-Specker
and the Greenberger-Horne-Zeilinger theorems, as for the latter ones
any violations of the conditions of possible experience must occur
on every single quantum and at least for a single observable~\cite{svozil_2010-pc09}
rather than occasionally.

Those results relate to situations in which {\em omniscience} is assumed; that is,
all observables which could potentially be observed can indeed
be associated with actual elements of physical reality of a single quantum.
For a realist this might appear self-evident~\cite{stace1}.
Also for experimentalists this seems to be obvious;
after all, any particular observation renders outcomes,
regardless of the mutual complementarity of some of the observables involved;
in this view, ``potentially operational'' means  ``existence.''
By this inkling, the situation suggests that the measurement ``reveals''
a pre-existing element of physical reality of the quantum observed.
Stated pointedly,
registration of some detector clicks is interpreted as a revelation
about what is taken as the quantized object.

If these pre-existing elements of physical reality are taken for granted,
it is not unreasonable to ``solve''
or ``explain'' the conundrum imposed by the various aforementioned theorems
by assuming that
any potential measurement outcome
may depend on whatever other maximal co-measurable collection of observables
(the context, interpretable as maximal operator~\cite[sect.~84]{halmos-vs})
are co-measured alongside.
This dependency of the outcome of a single quantum measurement on its context
is termed {\em contextuality}.

Note that the Born rule, and also Gleason's theorem,
requires the quantum probabilities and expectations, and thus
all quantum statistical properties, to be {\em noncontextual.}
Notice also that
contextuality attempts to maintain a realistic, omniscient,
quasi-classical framework by abandoning context independence for single quantum observables.

Now, if one maintains realistic omniscience --
that is, the pre-existence of all outcomes of complementary potential observables --  then it is indeed true that,
as stated by Cabello~\cite{cabello:210401},
``the immense majority of the experimental violations of Bell inequalities [[proves]] quantum contextuality.''
Actually, the only difference between older evidence of violations of Bell-type inequalities
and more recent ones~\cite{hasegawa:230401,Bartosik-09,kirch-09,Lapkiewicz-11}
seems to be based on the fact that the prior ones rely on spatially separated
quanta in Einstein-Podolsky-Rosen ``explosion'' type schemes, whereas more recent ones are based on single quanta --
a concept which appears to be more in the spirit of Kochen-Specker type theorems which
apply to the structure of observables of single quanta~\cite{cabello-nature-2011}.
But even these sorts of empirical findings referring to single quanta rely on the
non-instantaneous measurement of all but a few (mostly two or three in cases involving two- or three-particle
Einstein-Podolsky-Rosen and Greenberger-Horne-Zeilinger type) configurations,
and therefore cannot even counterfactually assure the operational existence of all elements of physical reality at once~\cite{svozil-2006-uniquenessprinciple}.

Alas, these assumptions are neither necessary (and sufficient),
as other, rather exotic options~\cite{pitowsky-82,meyer:99} demonstrate,
nor is there any more direct empirical evidence in their support.
Indeed, quantum predictions of Einstein-Podolsky-Rosen type setups involving
singlet states of qutrits suggest that contextuality
cannot be observed~\cite{svozil:040102}, although a direct experimental test is still lacking.

Thus with regards to quantum contextuality, the situation is rather discomforting:
insofar as contextuality seems to ``explain''
various findings related to quantum predictions and correlations,
it can only be indirectly inferred by assuming some extra assumptions, including classical omniscience;
otherwise it is not necessary.
And insofar it could be directly testable it is very unlikely to show up.
Because of this dilemma,
it is suggested to re-evaluate recent empirical findings in terms of a much broader picture
of {\em value indefiniteness;} including also the possibility
that there needs not exist a pre-existing element of physical reality
associated with certain observables.

In certain situations the experimental outcomes actually measured might not originate from such pre-existence,
but might depend on the interaction
between the quantum measured and the measurement apparatus.
Pointedly stated, the outcome might not reflect an intrinsic objective physical property
of the quantized object,
but rather originate in the way a measurement apparatus generates the outcome by interacting with the quantum.
Already Bell~\cite{bell-66} suggested that
``the result of an observation may reasonably depend $\ldots$
on the complete disposition of the apparatus.''
Perhaps this was also what Bohr had in mind by mentioning~\cite{bohr-1949}
``the impossibility of any sharp separation between the behaviour of atomic
objects and the interaction with the measuring instruments which serve to define the conditions
under which the phenomena appear.''

Acknowledgements: This research was partly supported by Seventh Framework Program for research and technological development (FP7), PIRSES-2010-269151-RANPHYS.

%\bibliography{svozil}


\end{document}